# OBSERVATION OF $\nu_\tau$ CHARGED-CURRENT INTERACTIONS

J.SIELAFF for the DONUT collaboration
*University of Minnesota, Minneapolis MN 55455, USA*

The DONUT collaboration has observed 4 events in a hybrid emulsion spectrometer detector consistent with $\nu_\tau$ charged-current interactions. The resulting tau lepton is identified by its decay into a single charged particle. The expected background of charm decays and secondary interactions is .34±.05 events. The probability that all $\nu_\tau$ candidates are due to background processes = $1.1 \times 10^{-4}$

1. Introduction

203 neutrino interactions observed with the DONUT detector were analyzed for the presence of tau neutrino charged-current interaction as evidenced by the presence of a short lived tau lepton. 86.4% [1] of tau decays result in a single charged particle and are identified as a kink along the track recorded in the emulsion. The superb resolution of the emulsion (3mr angle and .4micron position for an emulsion track segment) allows kinks down to ~10mr to be observed. Only kink events having no identified electron or muon from the interaction vertex are considered.

2. DONUT neutrino beam and detector

The neutrino beam enriched with tau neutrinos was produced by directing 800 GeV protons from Fermilab's Tevatron onto a one-meter long tungsten dump. All flavors of neutrinos where then produced from primarily the decays of the charmed mesons. Tau neutrinos are produced from $D_s$ decays which result in a neutrino anti-neutrino pair through the decay chain: $D_s \to \tau + \bar{\nu}_\tau$, $\tau \to \nu_\tau + X$. The flux of neutrinos and anti-neutrinos from these sources was $(\nu_e\ \nu_\mu\ \nu_\tau) = (4.6 \pm .8,\ 4.4 \pm 1.0,\ .79 \pm .13) \times 10^4$/interacting proton [2]. Non-prompt $\nu_\mu$ are also produced from light mesons that decay before they interact in the dump and account for 44%[3] of the $\nu_\mu$ interactions.

The emulsion detector is located 36 m downstream of the tungsten beam dump and is shielded from charged particles from the dump by 2 large magnets and several meters of concrete and steel. Several different styles of emulsion detectors were used in the experiment to test their effectiveness for locating tau decays. The emulsion detector was modular, consisting of several stacks of 50 cm x 50 cm emulsion sheets oriented perpendicular to the beam. The sheets had either 100 or 350 μm of nuclear emulsion on either side of a supporting plastic base. In five of the seven modules 1 mm thick steel sheets were layered between the emulsion sheets to provide the mass necessary for neutrino interaction. The average mass of the emulsion stacks was 261 kg with an exposure of $3.6 \times 10^{17}$ protons on the dump. Emulsion stacks with steel accounted for 70% of the exposed mass.

A conventional spectrometer consisting of a scintillation counter trigger, an analysis magnet, drift chambers, a lead and scintillating glass calorimeter and a muon identification system were used to reconstruct the event and aid in primary lepton identification. A scintillating fiber tracker with tracking planes between the emulsion modules was used to pinpoint areas in the emulsion to scan.

3. Neutrino event reconstruction and tau decay search

After exposure the emulsion stacks were developed and regions predicted to contain a neutrino interaction vertex were scanned by an automated system.[5][6] Particle tracks were reconstructed by linking segments in successive emulsion layers and neutrino interactions were identified by a vertex of two or more tracks. 203 located neutrino interactions are used for this tau decay search.

Primary tracks were linked to hits in the downstream detectors to reconstruct the neutrino interaction. The momentum of each track making up an interaction vertex was measured be either multiple scattering within the emulsion stacks or by bending in the analysis magnet. Primary electrons were identified from pair production in the emulsion or the shower development in the scintillating fiber tracker and primary muons were flagged with the muon identification hodoscope.

All tracks in the 203 events were searched for a tau decay like kink with the following criteria: at least one segment of the parent track visible in the emulsion, impact parameter of the daughter tack to the interaction vertex < 500 μm, the kink angle greater than 10 mr and less than 250 mr, length of parent track < 5 mm, daughter momentum > 1 GeV/c and

$P_t$ of the kink $\geq 250$ MeV/c. When applied to simulated tau neutrino interactions theses cuts retained 50% of the tau decays. Events selected which had an identified primary electron or muons were rejected.

Four events were found which satisfied the above criteria. Figure 1. shows the tracks reconstructed from the emulsion segments to form a neutrino interaction and tau decay.

Figure 1: Emulsion segments of the four events surviving tau decay cuts. Clockwise from upper-left: 3024-30175 τ→electron; 3039-01919 τ→hadron, 3263-25105 τ→hadron, 3333-17665 τ→electron. Target material is shown on the bar along the bottom: plastic (no shading) emulsion (hatched) and steel (shaded). This figure is taken from reference 4.

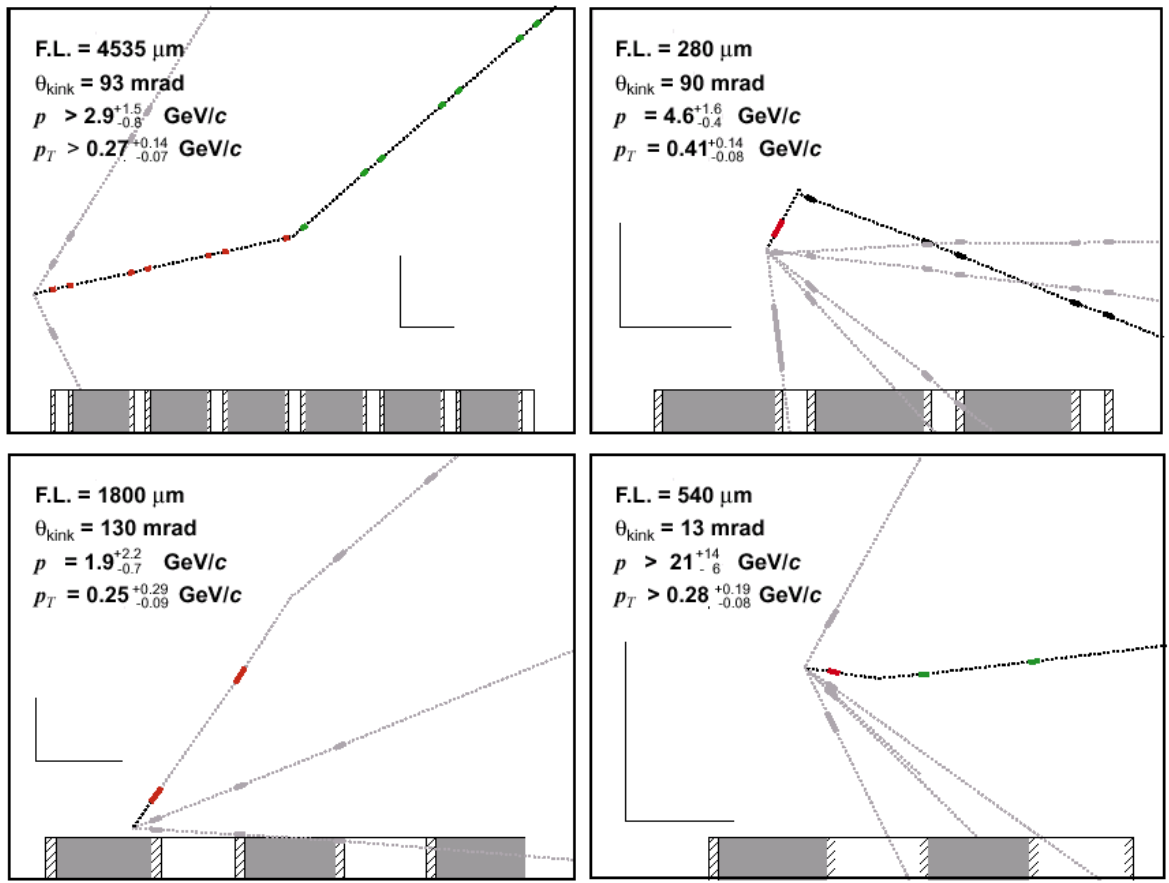

4. Backgrounds to the tau signal

Backgrounds to the tau signal are due kinks found in neutral-current interactions or charged-current interactions in which the primary electron or muon is not identified. Sources of kinks in these events include secondary interactions of primary particles and primary charm decays.

A scatter or inelastic interaction in which only one produced particle is detected in the emulsion may mimic tau decay. Simulations of hadrons produced in neutrino interactions with the energy spectrum and composition of DONUT's neutrino beam and interacting in the emulsion detector predict a background of .18 ± .02 with $P_t$ greater than 250 MeV/c for 203 interactions.

Charged charm produced in charged-current interactions that decay to a single charge particle may also be a background to the tau signal. An estimate of 7.7% of the 168 ± 16 charged-current interactions in the data set of 203 events produce charm. Accounting for production of charged charm, branching ratios to single-charge decay, lepton identification efficiency and the cuts used in tau candidate selection, .16 ± .04 charm background events are expected.

A Bayesian analysis of each selected event was used to calculate its probability of being due to tau neutrino interaction or a background source. The parameters used in the likelihood functions were: polar angle balance between the tau track and all other charged tracks from the neutrino interaction vertex, production angle between the tau and the neutrino, decay length and kink $P_t$. The prior probability functions depend on daughter type and the material in which the kink occurs. Results of this analysis are summarized in table 1.

From this analysis the probability that all four observed events are due to background sources is $1.1 \times 10^{-4}$

Table 1: Results of Bayesian analysis of the four neutrino events selected with tau cuts. Probabilities of tau decay, charm and secondary interaction backgrounds are shown.

|  | 3039-01910 | 3024-30175 | 3263-25102 | 3333-17665 |
| --- | --- | --- | --- | --- |
| Tau decay | .97 | .88 | .03 | .97 |
| Charm decay | .03 | .12 | .02 | .03 |
| $2^{ndry}$ interaction | $1.4 \times 10^{-4}$ | $<10^{-5}$ | .95 | $<10^{-5}$ |